\documentclass[aps,prl,twocolumn,amsmath,amssymb,floatfix]{revtex4-1}
\usepackage{tabularx}
\usepackage{bm}
\usepackage{euscript}
\usepackage{epsfig,psfrag,subfigure}
\usepackage{graphicx}
\usepackage{color}
\usepackage{amsfonts}
\usepackage{exscale}
\usepackage{amsbsy}

\def\avg#1{\left\langle#1\right\rangle}

\def\be{\begin{equation}}       \def\ee{\end{equation}}
\def\bea{\begin{eqnarray}}      \def\eea{\end{eqnarray}}
\def\ba{\begin{array} }
\def\ea{\end{array} }
\def\bnum{\begin{enumerate} }
\def\enum{\end{enumerate}}

\def\nn{\nonumber}

\def\=>{\Rightarrow}
\def\>{\rightarrow}
\def\A{\uparrow}
\def\V{\downarrow}

\def\eye2{Fathbb{I}}

\def\Eq#1{Eq.~(\ref{#1})}
\def\Fig#1{Fig.~\ref{#1}}

\renewcommand{\v}[1]{{\bf #1}}

\newcommand{\s}{{\sigma}}

\newcommand{\Ref}[1]{Ref.~\cite{#1}}

\renewcommand{\>}{\rangle}

\newcommand{\ga}{\gamma}

\begin{document}

\title{Charge-4e superconductors: a Majorana quantum Monte Carlo study}
\author{Yi-Fan Jiang$^{1,2}$, Zi-Xiang Li$^1$, Steven A. Kivelson$^2$, and Hong Yao$^1$}
\affiliation{$^1$Institute for Advanced Study, Tsinghua University, Beijing 100084, China\\
$^2$Department of Physics, Stanford University, Stanford, CA 94305, USA}

\begin{abstract}
\end{abstract}

\date{\today}
\maketitle

{\bf Many features of charge-4e superconductors remain unknown because even the ``mean-field Hamiltonian'' describing them is an interacting model. Here we introduce an interacting model to describe a charge-4e superconductor (SC) deep in the superconducting phase, and explore its properties using quantum Monte Carlo (QMC) simulations. The QMC is sign-problem-free, but only when a Majorana representation is employed. As a function of the chemical potential we observe two sharply-distinct behaviors:  a ``strong'' quarteting phase in which charge-4e quartets are tightly bound (like molecules) so that  charge-2e pairing does not occur even in the temperature $T\to 0$ limit, and a ``weak'' quarteting phase in which a further transition to a charge-2e superconducting phase occurs at a lower critical temperature. Analogous issues arise in a putative $Z_4$ spin-liquid with a pseudo-Fermi surface and in characterizing charge-2e SCs with ``odd frequency pairing'' (or equivalently with a ``composite order parameter'') and metallic spin-nematics. Under certain circumstances, we also identified a stable $T=0$ charge-4e SC phase with gapless nodal quasiparticles. We further discuss possible relevance of our results to various  experimental observations in 1/8-doped LBCO.}

Superconductors, ranging from weakly correlated systems such as Hg to strongly correlated ones including the high-$T_c$ cuprates and Fe-based superconductors, have been among central focuses of physics research for over a century\cite{Schriefferbook, XGWenbook, Fradkinbook}. Heuristically, superconductivity occurs when electrons bind into bosonic pairs which then condense to form a phase coherent quantum fluid. The order parameters describing all known superconductors carry charge-2e.  Nonetheless, condensing charge-2e Cooper pairs is not the only possible way to achieve superconductivity. One intriguing possibility which is beyond the scope of conventional BCS theory is a charge-4e superconducting state, in which the condensate carries charge-4e while pair-field correlations vanish at long distances. The flux quantum in such a phase is  $\phi^*=hc/4e$ and the current oscillations in a DC biased Josephson junction oscillate with a frequency $\omega = 4eV/\hbar$.

To date, the existence of a charge-4e superconducting state has not been established in any material. While charge-4e ``quartet-field'' operators are long-range correlated even in an ordinary charge-2e SC, these correlations are simply a harmonic of the fundamental condensate.  However, a charge-4e superconductor may be realized as a ``partially melted'' phase intermediate between a charge-2e SC and a metal;  here strong fluctuations destroy the long-range charge-2e order, leaving the charge-4e order behind as ``vestigial'' long-range order\cite{Nie2014pnas}. Such a scenario was explored by Berg, Fradkin, and Kivelson in studying how putative pair-density-wave order in certain high-temperature SCs\cite{Tranquada2007prl,Berg2007prl} is melted thermally\cite{Erez2009,Erez2009a}. A similar scenario was considered by Radzihovsky and Vishwanath\cite{Leo2009prl} in the context of the multistep disordering of an FFLO state. Other ways to achieve charge-4e superconductivity include interactions that favor quarteting rather than pairing\cite{SteveVic} and condensing charge-4e skyrmions in quadratic-band-touching systems\cite{Senthil2009,KSun2009,Moon2012}.

\begin{figure}[t]
\centering
\includegraphics[width=5cm]{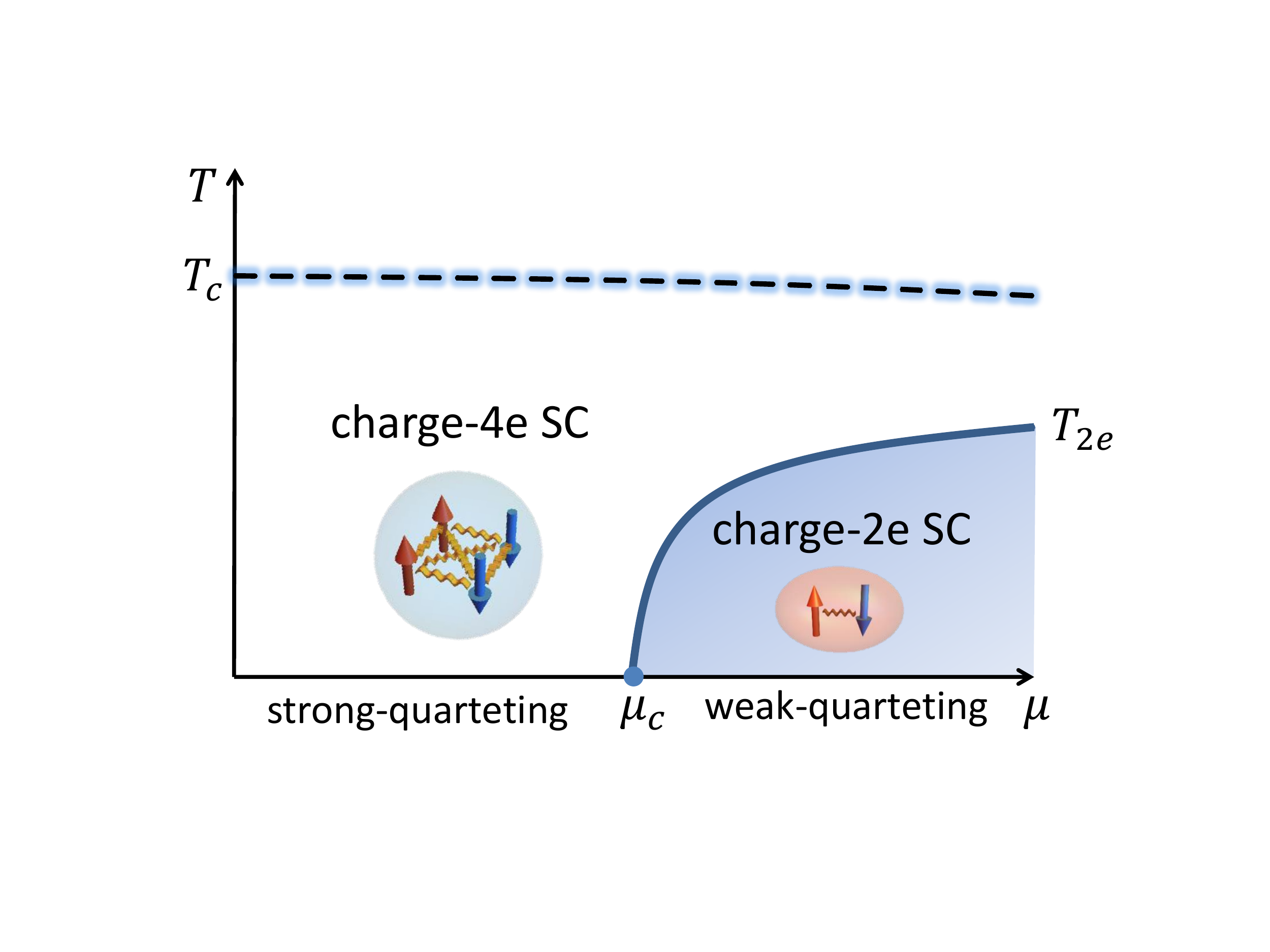}
\caption{Schematic phase diagram of a system with a high-temperature transition, $T_c$, (the dashed line)  to a charge-4e superconducting state.  A  ``strong-quarteting'' regime appears for values of the chemical potential, $\mu$, less than a critical value, $\mu_c$, where charge-4e quartets are tightly-bound so that charge-2e pairing cannot occur even as $T\to 0$. A crossover occurs at $\mu_c$ to a ``weak'' quarteting regime such that for $\mu>\mu_c$  a second (Ising) transition occurs at $T_{2e}>0$ to a charge-2e superconductor. Note that the minimal model we studied applies only much below the dashed line.}
\label{phasediag}
\end{figure}

Many physical properties of a charge-4e SC are unknown even in principle due to the fact that even the ``mean-field Hamiltonian'' describing such a phase must itself be an interacting model. This is in sharp contrast with charge-2e SCs whose mean-field BCS Hamiltonian is quadratic in fermionic operators so that it can be solved exactly. For charge-4e superconductors, there are a number of open questions. For instance, one naturally asks whether a charge-4e SC is stable at zero temperature\cite{Barkeshli2013}.

As with charge-2e superconductors, charge-4e superconductors can be distinguished according to their transformation properties under the space-time symmetries of the system -- $s$-wave, $d$-wave, $p+ip$-wave, etc.; here we will consider only the $s$-wave case.  In addition, there exist two sharply-distinct types of quarteting phases: a ``strong'' quarteting phase where charge-4e quartets are tight-bounded (like molecules) and charge-2e pairing cannot occur even at zero temperature, and a ``weak'' quarteting phase where quartets are loosely bound so that charge-2e pairing can emerge at a lower critical temperature. A sharply defined quantum critical point at $T=0$ separates a regime in which the strong-quarteting phase is stable all the way to $T=0$, from a weak-quarteting phase where a further transition to a charge-2e SC occurs at $T_c>0$, as schematically shown in \Fig{phasediag}.

Here we introduce and study a minimal interacting model [see \Eq{H4e} below] to describe charge-4e superconductors at the mean-field level in which the charge conservation symmetry is explicitly broken by the presence of a charge-4e quarteting field that plays a role analogous to that of the pairing field in a charge-2e superconductor. Because even the mean-field model is interacting, inferring its properties is still highly non-trivial.  Remarkably, the model is sign-problem-free in quantum Monte Carlo simulations\cite{Scalapino1981a, Hirsch1981, Sugiyama86, Sorella1992,Loh1990,Troyer2005,Berg2012} only when the Majorana representation\cite{ZXLi2015} is employed.

By performing large-scale Majorana quantum Monte Carlo (MQMC) simulations\cite{ZXLi2015}, we explore the thermodynamic and spectral properties of a putative charge-4e SC at temperatures well below $T_c$.  Among the questions we address are whether the charge-4e SC is gapped or gapless, and whether there is a further transition at $T_{2e} < T_c$ to a charge-2e SC which spontaneously breaks the $\mathbb{Z}_4$ charge conservation to $\mathbb{Z}_2$\cite{Barkeshli2013}.

{\bf The model:}
The following mean-field model is introduced to describe a charge-4e SC well below $T_c$:
\bea
H&=&-t\sum_{\avg{ij},\s}\big[c_{i\s}^\dagger c_{j\s}+H.c.\big] - \mu\sum_{i,\s} c_{i\s}^\dagger c_{i\s}\nn\\
&&~~~~~~~~~~+ V \sum_{\avg{ij}}\big[c_{i\A}^\dagger c_{i\V}^\dagger c_{j\V}^\dagger c_{j\A}^\dagger+H.c.\big] ,\label{H4e}
\eea
where $t>0$ is the nearest-neighbor hopping integral, $\mu$ the chemical potential, and $c_{i\s}$ the electron annihilation operator at site $i$ with spin polarization $\s=\uparrow$/$\downarrow$. The interaction term in \Eq{H4e} describes the charge-4e quarteting on nearest-neighbor bonds and $V$ thus represents the mean field corresponding to a singlet charge-4e condensate.  Thus, $V$ vanishes at all $T > T_c$, and can be approximated by a $T$ independent constant $V \sim T_c$ for $T \ll T_c$.  As required, this model explicitly breaks the charge $U(1)$ symmetry to $Z_4$, namely electron number is conserved only modulo 4. We take $V$ to be positive as its phase can be changed by gauge transformation $c_{i\s} \rightarrow e^{i \theta} c_{i\s}$ while keeping other terms in \Eq{H4e} invariant.

{\bf Mean-field solution:}  We can make a first pass at studying the BCS instability of the charge-4e state by applying mean field theory to the mean-field Hamiltonian in \Eq{H4e}.  Using the obvious mean-field decoupling of the interacting term results in two gap parameters:  $\Delta_0 = V \langle c_{i\uparrow}^\dagger c_{i\downarrow}^\dagger\rangle_{bcs}$ and $\Delta_1 = V \langle [c_{i\uparrow}^\dagger c_{j\downarrow}^\dagger+ c_{j\uparrow}^\dagger c_{i\downarrow}^\dagger]\rangle_{bcs}$ where $i$ and $j$ are any pair of nearest-neighbor sites and $\langle \ \rangle_{bcs}$ is the thermal average with respect to the corresponding quadratic BCS Hamiltonian.  If we were to work at fixed average electron density per site, $n$,  in addition to the self-consistency equations for $\Delta_0$ and $\Delta_1$, we would need to compute the chemical potential, $\mu$, self-consistently, as well.  To make easier contact with our Monte Carlo results, we will instead discuss the results at fixed $\mu$ and compute $n(T,\mu,V)$ when desired.

The resulting mean-field phase diagram is qualitatively reproduced in Fig. 1.  The nature of these results are familiar from the BCS theory of charge-2e SCs.   We will focus the discussion on what is plausibly the most interesting regime of parameters, $T_c \ll 8t $ (where $8t$ is the band-width) and $V/8t \ll 1$. The spectrum of quasiparticle excitations is
\be
E(\vec k) = \sqrt{[t\gamma_{\vec k} + \mu]^2 + [\Delta_0+ \Delta_1\gamma_{\vec k}]^2}
\ee
where $\gamma_{\vec k} = 2(\cos k_x + \cos k_y)$ is the nearest-neighbor structure factor.
The critical value of $\mu_c\to -4t$ in the limit $V\to 0$;  here, for $\mu < \mu_c$ the density of particles $n\to 0$ as $T\to 0$.  However, for non-zero $V$, the electron density is a smooth function of $\mu$ such that $n(0,\mu, V) \sim 4 [V/4|\mu_c - \mu|]^2$ for $\mu_c - \mu \gg V$, where $4|\mu_c - \mu|$ represents the typical kinetic energy cost to create a charge-4e quartet. Moreover, we can show that $\mu_c = -4t - a V^2/t + \ldots$ in the small $V$ limit, where $a$ is a constant of order 1. For $\mu > \mu_c$, the critical temperature $T_{2e} \sim [\mu-\mu_c] \exp[ - 1/4N(\mu)V]$ where $N(\epsilon)$ is the density of states of the non-interacting problem.  Notice that so long as $V/t \ll 1$,  $T_{2e} \ll T_c$, ignoring the $T$ dependence of $V$ is self-consistently justified.

Several features of this approximate solution warrant mention.  In the strong quarteting limit, where $\mu < \mu_c$, the quasiparticle spectrum is gapped even in the absence of charge-2e condensation, $\Delta_a=0$.  Conversely in the region of weak quarteting, there is a ``pseudo-Fermi-surface'' of gapless quasiparticle modes for $T>T_{2e}$.  Since $T_{2e}$ is exponentially small, this pseudo-Fermi surface is well defined, even though it only exists at non-zero temperatures. However, the quasiparticles are not the usual ones of a Fermi liquid, in that their charge is only conserved mod 4. We shall see that salient features of this ``double-mean-field'' analysis are reproduced by our QMC simulations.

\begin{figure}[t]
\centering
\subfigure{\includegraphics[height=2.8cm]{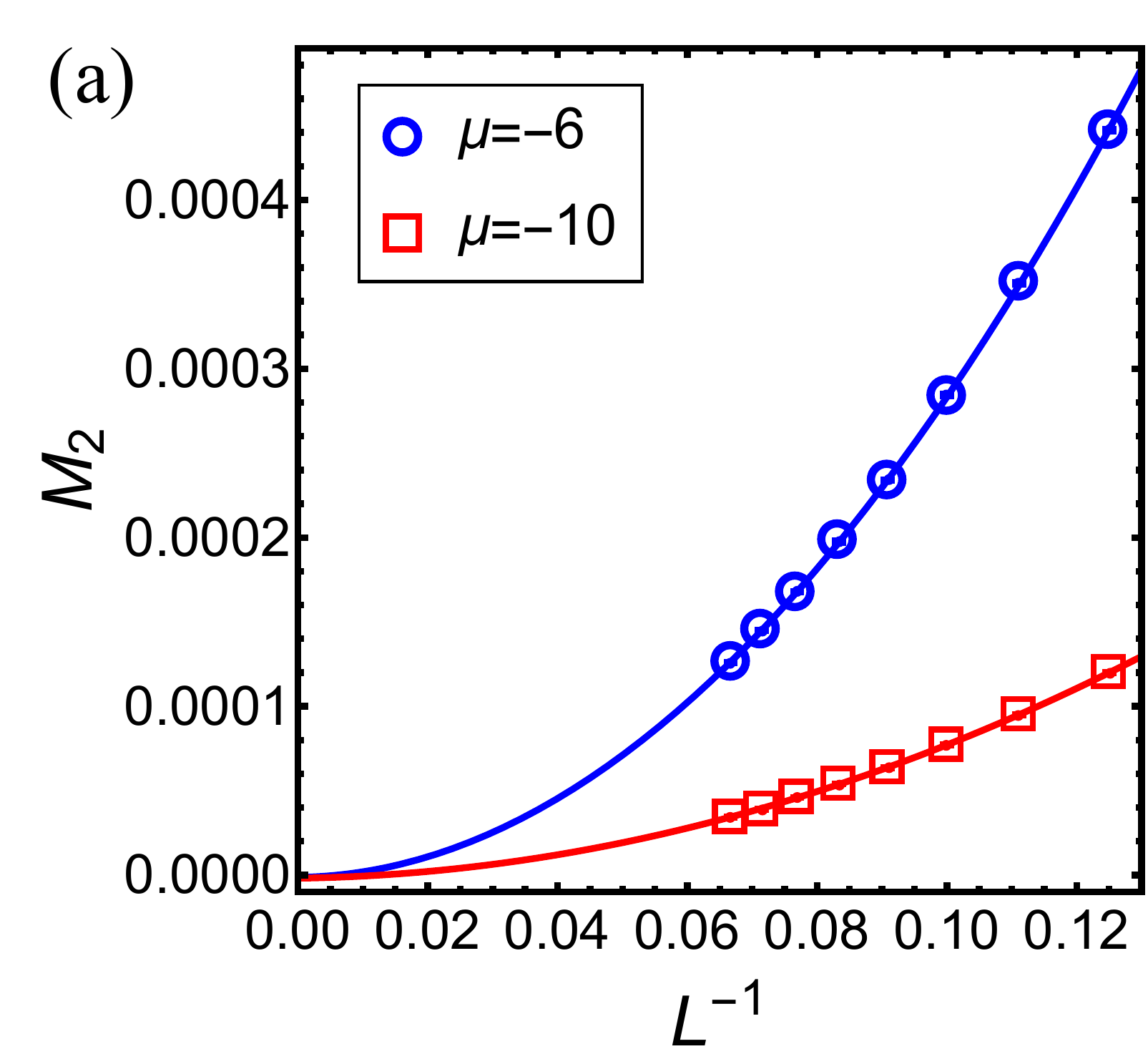}}~
\subfigure{\includegraphics[height=2.8cm]{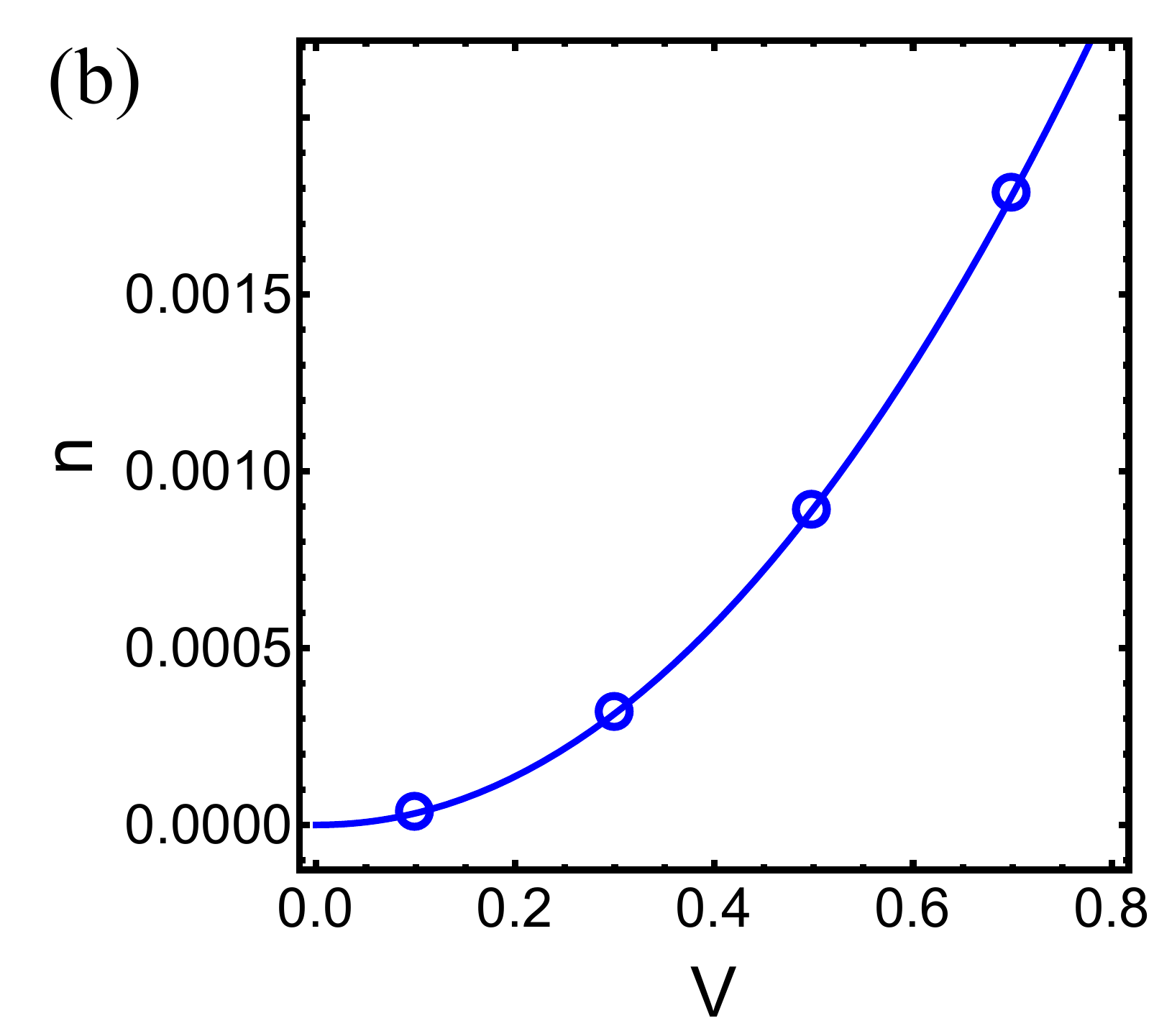}}
\subfigure{\includegraphics[height=2.8cm]{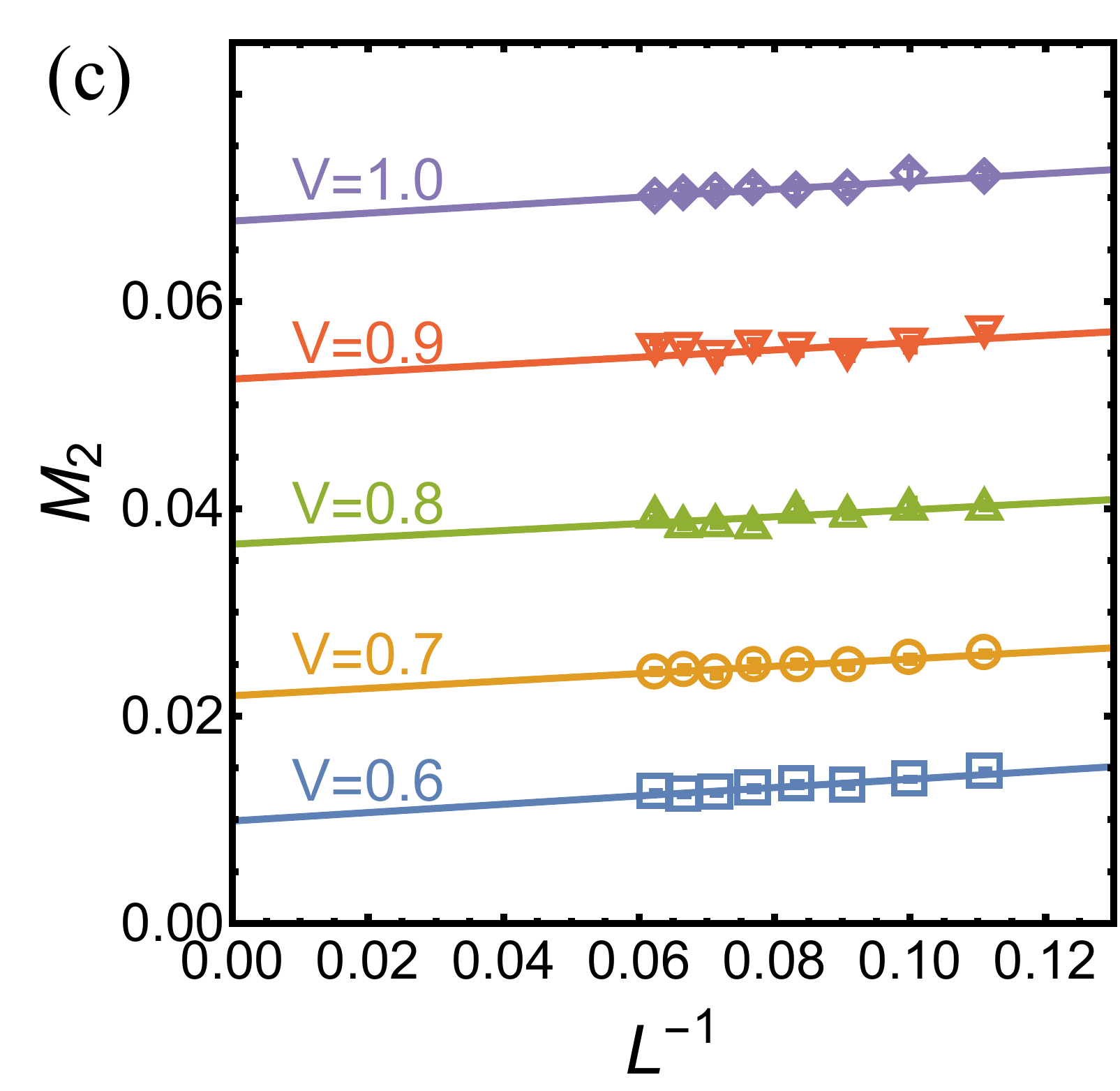}}~~~~
\subfigure{\includegraphics[height=2.8cm]{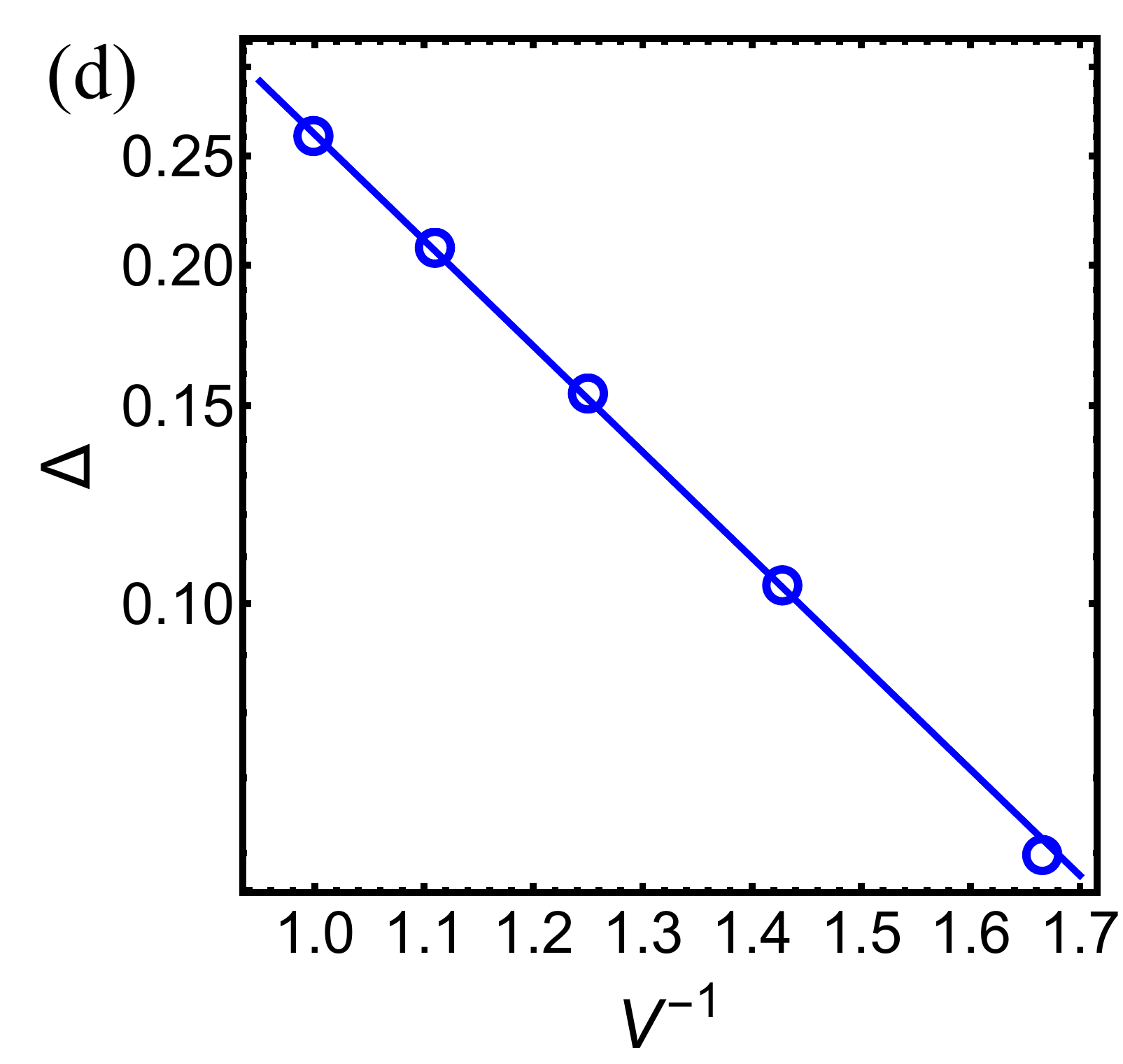}}
\caption{(a) The finite-size scaling of the charge-2e pairing structure-factor $M_2$ on the square lattice of $V=1.0t$ for $\mu=-6.0t$ and $-10.0t$, respectively (in the ``strong-quarteting'' region) with size $L=9,10,\cdots,15$. (b) The quadratic polynomial fitting of particle density $n(L\to\infty)$ for $\mu=-10.0t$, with $V$ from $0.1t$ to $0.7t$;  the solid line fitted  through the data is $n=0.0037\ (V/t)^2$. Here, we obtain $n(L\to\infty)$ from a finite-size scaling of $n(L)$ with $L=8,10,12,14$. (c) The finite-size scaling of $M_2$ for $V=0.6,0.7,0.8,0.9,1.0t$ and $\mu=-1.36t$ (finitely away from the half-filling). The system size $L$ varies from 9 to 16. Here, $M_2$ barely changes with increasing $L$. The colored lines are linear fits to  the data points for each $V$, which give rise to finite extracted values in the thermodynamic limit ($L\to \infty$). (d) The charge-2e pairing order parameter $\Delta\equiv V\langle c^\dag_{i\uparrow} c^\dag_{i\downarrow}\rangle=V[M_2(L\to \infty)]^{1/2}$ (plotted on a logarithmic scale) as a function of $V^{-1}$ for $\mu= -1.36t$ with $V/t$ in the range 0.6 to 1.0; the fit to the solid line shows that $\Delta\approx \alpha t e^{-gt/V}$ with $\alpha\approx 2.3$ and $g\approx 2.1$.}
\label{fin_mu}
\end{figure}

{\bf Majorana quantum Monte Carlo: }
We now proceed to study the zero-temperature properties of the model in \Eq{H4e} by performing QMC simulations. As there is no charge conservation in the charge-4e mean-field model, it is natural to employ the Majorana representation recently introduced in Ref. \cite{ZXLi2015} to solve the fermion-sign-problem in QMC simulations. We first introduce the Majorana representation of spin-1/2 electrons: $c_{i\s}=\frac1 2 (\ga_{i\s}^1+i\ga_{i\s}^2),\ c_{i\s}^\dagger=\frac1 2 (\ga_{i\s}^1-i\ga_{i\s}^2)$, where $\gamma^\tau_{i\sigma}$ are Majorana fermion operators with $\tau=1,2$ represents the Majorana indices and $\sigma=\uparrow,\downarrow$. The quarteting interaction in the Majorana representation is:
$V(c_{i\A}^\dagger c_{i\V}^\dagger c_{j\V}^\dagger c_{j\A}^\dagger\!+\!h.c.)=\frac{V}{32}\sum_{\alpha=1}^4 \Big[i \ga_i^t B_\alpha \ga_j\Big]^2$,
where $\ga_i^t\equiv (\ga_{i\A}^1, \ga_{i\A}^2, \ga_{i\V}^1, \ga_{i\V}^2)$, $B_1=\s^z\tau^z$, $B_2=i\s^0\tau^z$, $B_3=\s^0\tau^x$, and $B_4=i\s^z\tau^x$ ($\s^i$ and $\tau^i$ are the Pauli matrices acting in the spin and Majorana space, respectively). Upon application of the Trotter decomposition and Hubbard-Stratonovich transformations, the decoupled Hamiltonian at imaginary time $\tau$ is
\bea
\hat{h}=\sum_{\langle ij\rangle} \Big[\frac{-t}{2}\ga_i^t\s^0\tau^y \ga_j \!+\!\frac{V}{32}  \eta_{ij}^\alpha \ga_i^t B_\alpha \ga_j \Big]\!-\!\frac{\mu}{4}\sum_i \ga_i^t\s^0\tau^y \ga_i,~~~
\eea
where $\eta_{ij}^{\alpha}$ are imaginary time dependent auxiliary fields on bonds $\avg{ij}$, and the summation over $\alpha=1,2,3,4$ is implicit.  It is straightforward to check that $\hat h$ possesses two anti-commuting Majorana-time-reversal symmetries: $T^- = i \s^y \tau^x K$ and $T^+=\s^x\tau^x K$. According to the general classification scheme of the fermion-sign-problem proposed in \Ref{ZXLi2016}, the Majorana-bilinear operators respecting both $T^+$ and $T^-$ belong to the recently-introduced sign-problem-free Majorana-class\cite{ZXLi2015}, which is one of two fundamental sign-problem-free symmetry classes\cite{ZXLi2016} (the other one is ``Kramers-class''\cite{CWu2005}).

{\bf Strong-quarteting phase: }
The strong-quarteting phase is realized when $\mu<\mu_c$; here there is a gap in the spectrum at $V=0$, so a pairing instability should not occur for weak $V$ because it is unlikely for the system to develop a charge-2e pairing in the absence of Fermi surfaces. We first perform projector MQMC simulation to study the zero-temperature properties of the charge-4e model on the square lattice with $\mu=-6t$ and $-10t$, respectively, both of which are much below the band bottom of $-4t$. We calculated the structure factor $M_2=\frac{1}{N^2}\sum_{ij}\langle c_{i\A}^\dagger c_{i\V}^\dagger c_{j\V}c_{j\A}\rangle$  of charge-2e $s$-wave pairing for $V=1.0t$ on lattices with size $L=9,10,\cdots,15$ ($N=L^2$). As shown in \Fig{fin_mu}(a), the finite size scaling of $M_2$ implies a vanishing charge-2e order parameter in the thermodynamic limit, as expected.

In the strong quarteting phase ($\mu<\mu_c$), the electron density $n$ remains finite as discussed in mean field section. We numerically study the electron density for $\mu=-10t$ and $V$ varied from $0.1t$ to $0.7t$ on the lattice with $L=8,9,\cdots,14$. As shown in \Fig{fin_mu}(b), after finite size scaling, the $n(L\to \infty)$ fits perfectly with $V^2$, as expected from the mean-field analysis.

{\bf Weak-quarteting phase: }
In the ``weak quarteting'' region ($\mu>\mu_c$), due to the finite density of states at Fermi surface, the quartet interactions are known to be marginally relevant \cite{Barkeshli2013}. Consequently, the system is inevitably unstable to charge-2e pairing at low enough temperatures which fully gaps the Fermi surface (or, in some circumstances, leaving discrete, gapless nodal points). So we perform MQMC simulations of the charge-4e models to investigate the possible charge-2e pairing for various $\mu$ above the band bottom.

For simplicity, we set $\mu=-1.36t$ for which the Fermi surface is relatively large. Even though we expect that charge-2e pairing should occur at any finite (even infinitesimal) value of $V$, finite-size effects are problematic for weak interactions because the charge-2e order parameter decreases exponentially as $V\to 0$. For this reason, we have taken values of $V=0.6,0.7,0.8,0.9$, and $1.0t$, which is relatively small compared with the band width, but not so small as to present calculation problems.

We compute the structure factor $M_2$ of the charge-2e $s$-wave pairing and then perform finite-size scaling analysis which clearly shows evidences of long-range pairing order, as shown in \Fig{fin_mu}(c). The charge-2e pairing amplitude $\Delta\equiv V\langle c^\dag_{i\uparrow} c^\dag_{i\downarrow}\rangle=V[M_2(L\to \infty)]^{1/2}$.  As the quarteting strength studied is relatively weak compared with the band width, we expect that $\Delta=V\langle c^\dag_{i\uparrow} c^\dag_{i\downarrow}\rangle$ is given by $\Delta \approx \alpha t\exp(-gt/V)$,
where $\alpha$ and $g$ are constants depending on the details of the Fermi surface. By fitting $\Delta$ obtained at $V=0.6\sim1.0t$ with the exponential form, our MQMC results show that $\alpha\approx2.3$ and $g\approx 2.1$, as shown in \Fig{fin_mu}(d). This supports that the charge-2e pairing occurs for any finite (even infinitesimal) quartet strength $V$, consistent with the RG analysis in Ref. \cite{Barkeshli2013}.

\begin{figure}[t]
\centering
\subfigure[]{\label{dirac}\includegraphics[height=2.3cm]{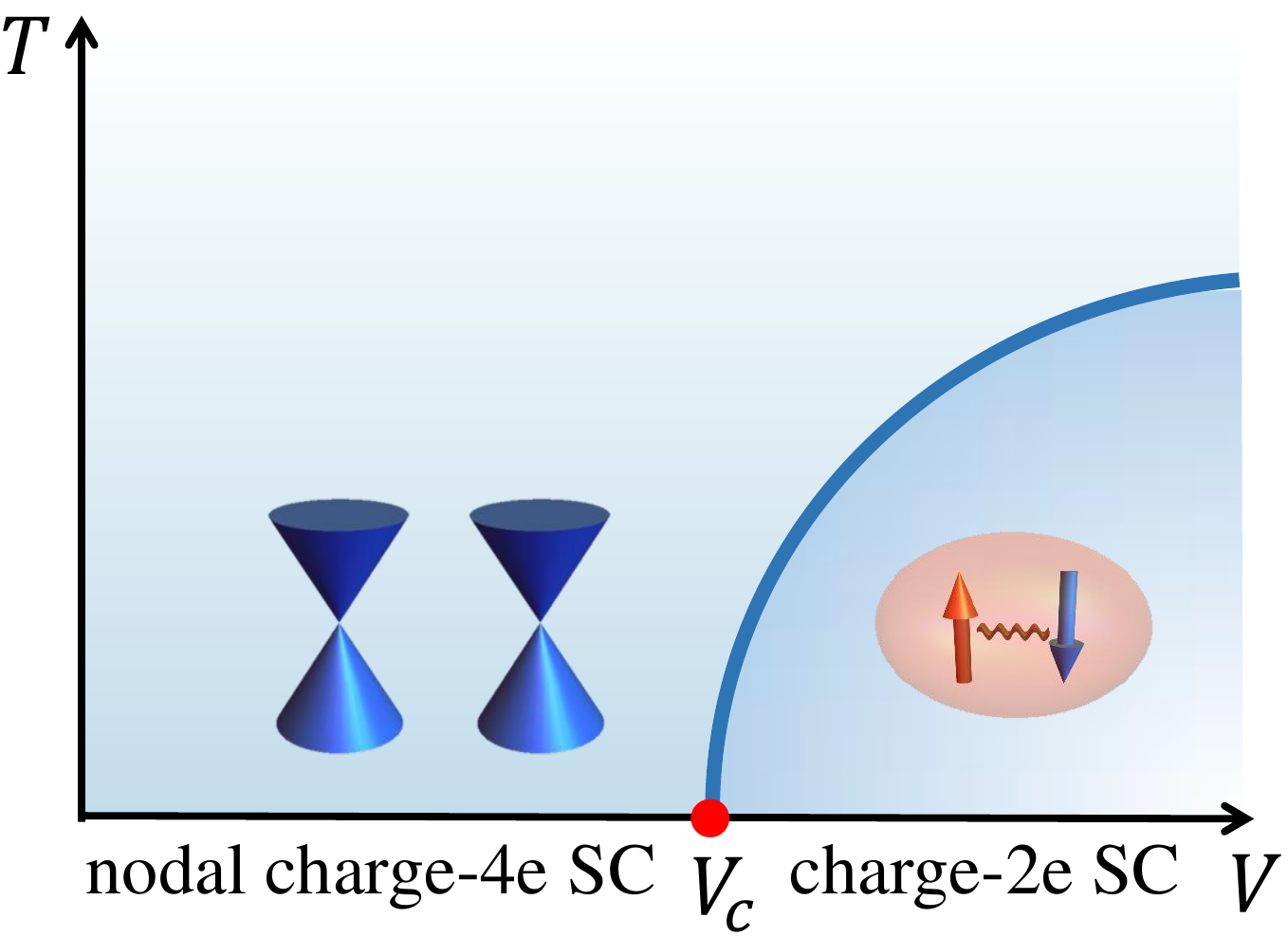}}
\subfigure[]{\label{vc}\includegraphics[height=2.32cm]{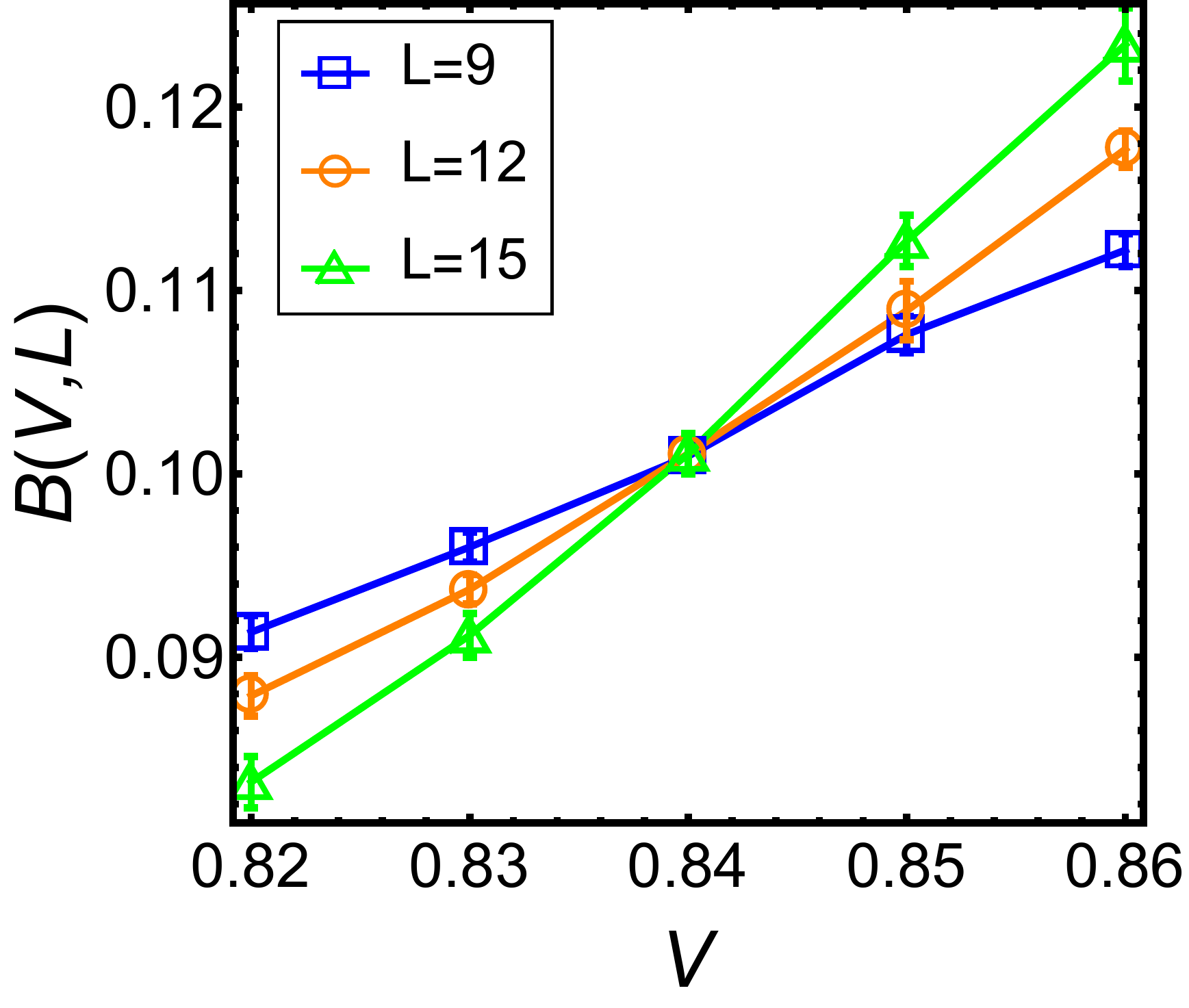}}
\subfigure[]{\label{expo:b}\includegraphics[height=2.32cm]{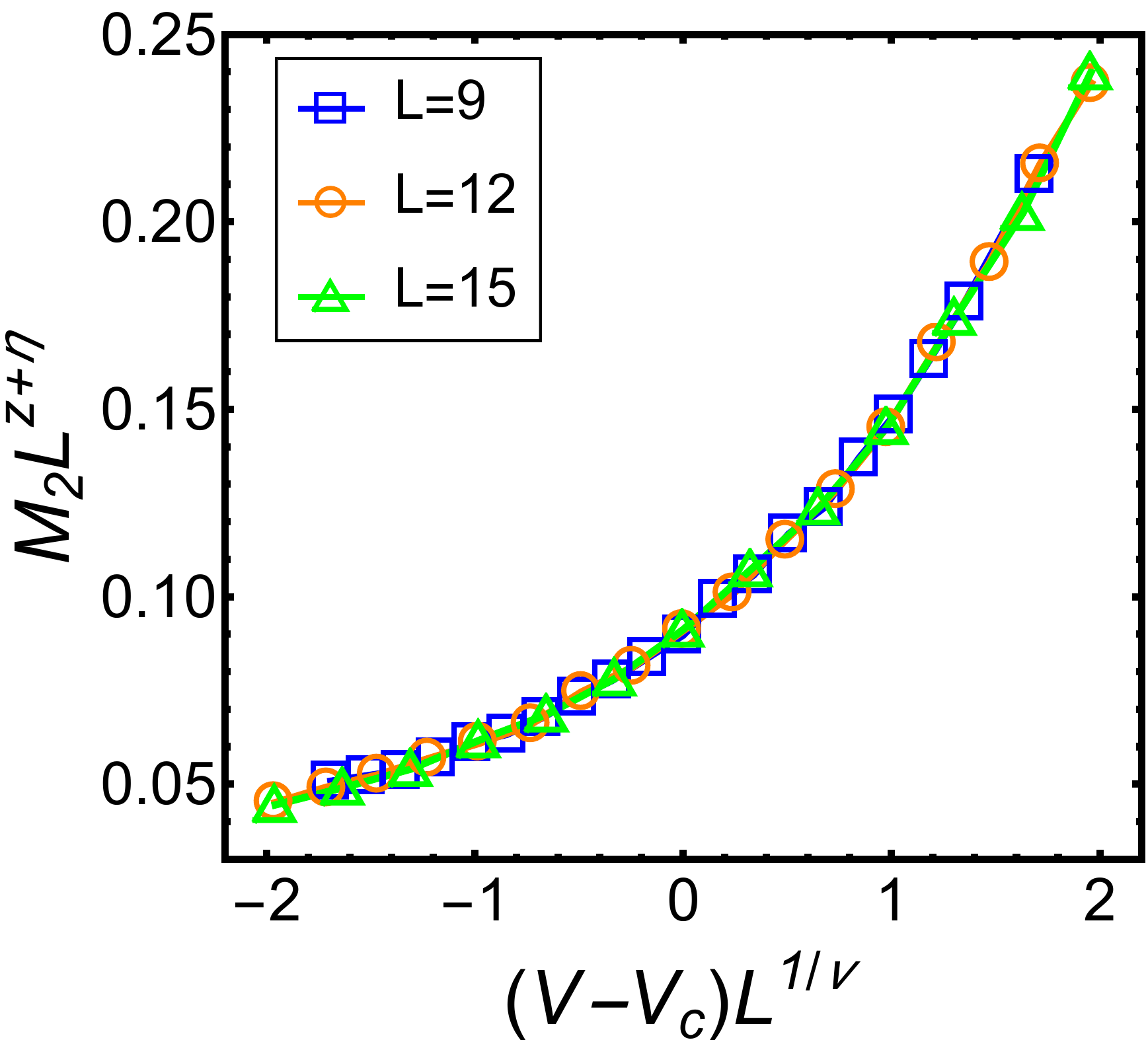}}
\caption{(a) A schematic phase diagram of the model on the honeycomb lattice at half filling ($\mu=0$). The quantum phase transition is expected to occur at a finite critical value of $V=V_c$, and to be in the Gross-Neveu universality class of Ising symmetry breaking. Correspondingly, for $V>V_c$, there should be a transition to a charge-2e SC with a finite temperature $T_c \sim (V-V_c)^{\nu}$ with $\nu\approx 0.78$ obtained from MQMC simulations.  (b) The Binder ratio $B(V,L)$ of the charge-2e pairing order parameter as a function of $V$ and $L$ for the case of half-filling. Here, $L$ varies from $9$ to $15$, as plotted in different colors. The crossing point gives rise to the critical point $V_c\approx 0.84t$. (c) The critical exponent $\eta\approx 0.52$ and $\nu\approx 0.78$ are obtained by reaching the best data collapsing. }
\end{figure}

{\bf Nodal charge-$4e$ SC:}
We expect that the two distinct phases we have identified on the square lattice and the nature of the transition between them are not sensitive to the details of the band structure. However, a distinct charge-4e SC phase can be shown to be stable, at least under the special circumstances in which the system supports a massless Dirac dispersion such as occurs on the honeycomb lattice at half-filling. We find that the charge-4e phase with a nodal quasiparticle is stable below a critical strength $V_c$, above which charge-2e pairing emerges in the ground state, as shown in \Fig{dirac}. Such a charge-4e to charge-2e quantum phase transition of spinful fermions belongs to the Gross-Neveu universality\cite{Gross1974, Rosenstein1993, Petersson1994, Rosa2001, Hofling2002, Herbut2009} in (2+1)D, similar to the ones studied previously \cite{Herbut2006, Raghu2008, Castro2009, Assaad2013, Lang2013, Castro2013, LWang2014,bZXL2015,Capponi2015,Pollmann2015, Daniel2015, SKJian2015, CenkeXu-15,ZXLi2015c}. For $V>V_c$, when the temperature is raised from zero, thermal fluctuation tends to destroy the charge-2e pairing such that the system undergoes a thermal phase transition to a charge-4e SC phase. The transition temperature $T_c\propto\Delta\sim (V-V_c)^{\nu}$ for $V$ close enough to $V_c$, with the critical exponent $\nu$ discussed below.

The MQMC simulations show that a uniform $s$-wave singlet pairing is the leading instability when $V$ is sufficiently large. To determine the critical value $V_c$, we calculate the modified Binder ratio $B(V,L)$\cite{Assaad2015} of the onsite $s$-wave pairing order parameter. For sufficiently large $L$, the Binder ratio $B$ should cross at $V=V_c$ for different $L$. The calculated $B(V,L)$ for various $V$ and $L$ are shown in \Fig{vc}, from which we find that the crossing occurs at $V_c\approx 0.84t$. The critical exponents $\eta$ and $\nu$ can be calculated from the scaling behavior of $M_2$ near the QCP:  $M_2 = L^{-1-\eta}\mathcal{F}(L^\frac{1}{\nu}(V-V_c))$. Here we implicitly assume the dynamical critical exponent $z=1$ since the low-energy theory of Dirac semimetals has an emergent Lorentz symmetry. Both $\eta$ and $\nu$ can be obtained by the data collapsing method: by choosing appropriate $\eta$ and $\nu$ we shall expect all the points $(L^{1/\nu}(V-V_c), M_2 L^{1+\eta})$ with various $V$ and $L$ collapsing into a single curve. In \Fig{expo:b}, we show that the single-curve collapsing is best for $\eta\approx 0.52$ and $\nu\approx0.78$. Because the charge-2e pairing spontaneously breaks the original $Z_4$ symmetry into $Z_2$ and massless Dirac fermions are gapped out by the Ising order parameter, this QCP should belong to the $N=2$ Gross-Neveu universality class of Ising symmetry breaking in (2+1)D. The results of $\eta\approx 0.52\pm 0.04$ and $\nu \approx 0.78\pm 0.06$ obtained by QMC are similar to the ones obtained in Ref. \cite{Herbut2009} by RG analysis using $\epsilon$-expansion up to the first order but somewhat smaller than the approximate results obtained from other RG analysis using large-$N$ or $\epsilon$-expansions \cite{Rosenstein1993, Petersson1994, Rosa2001, Hofling2002}.

{\bf Concluding remarks:}
We have not uncovered a region of parameters in which the charge-4e SC possess a pseudo-Fermi surface ({\it i.e.} a finite density of states for gapless quasiparticle excitations) which is stable as $T\to 0$.  However, it remains an open question whether addition of repulsive density-density interactions to Eq. \ref{H4e} could stabilize such a phase, as suggested in \cite{Barkeshli2013}.

There is a clear family resemblance between a charge-4e SC and other states which represent the condensation of four-fermion operators without any corresponding two fermion condensates.  Examples are a composite odd-frequency spin-singlet charge-2e SC with order parameter $\Delta_{comp} \equiv\vec M\cdot \vec \Delta$ where $\vec M$  and $\vec \Delta$ are the usual magnetization and spin-triplet charge-2e SC order parameters, and a spin-nematic order parameter, ${\cal N}_{ab} = M_aM_b - \vec M\cdot\vec M \delta_{ab}/3$.   Many of the same considerations that apply to the charge-4e SC have analogues in these phases, as well. In the context of spin-liquid physics\cite{Anderson,KRS},  various much discussed $Z_2$ spin-liquids\cite{Read-Sachdev-91,XGWen-91,Moessner-Sondhi-01} are formal analogues of distinct forms of charge-2e SCs.  It is thus natural to consider, as well, $Z_4$ spin liquids\cite{Barkeshli2013} which are analogues of charge-4e SCs. Finally, it is worth discussing the possible relevance of charge-4e SCs to the physics of certain special high $T_c$ superconducting cuprates. A number of spectacular, but at present still not conclusively understood thermodynamic and transport anomalies, have been documented in single crystals of  1/8-doped LBCO\cite{Tranquada2007prl}.  Below $T_{cdw} \approx 52K$ and $T_{sdw} \approx 42K$, charge- and spin-density wave (``stripe'') order, respectively, is clearly seen in diffraction experiments, with  long but finite correlation lengths (presumably due to quenched randomness).  The Meissner phase observed below $T_{meis} \approx 4K$  must reflect superconducting long-range order.  In the intermediate ranges of temperatures, the in-plane  and interplane resistivities appear to vanish below $T_{2d} \approx 17K$ and $T_{3d}\approx 10K$, respectively.  It has been suggested\cite{Berg2007prl} that the dynamical layer decoupling observed for $T_{2d}>T>T_{3d}$ may reflect the formation of pair-density wave (PDW) related phases, although because of the non-trivial interplay between  PDW order and quenched randomness,  even the phenomenological theory of these phases is incomplete.  It is plausible that in some range of temperatures, perhaps for $T_{meis} < T < T_{3d}$,  there may be vestigial charge-4e SC order, which only gives way to charge-2e SC order below $T_{meis}$.  It is important to note that, in this temperature range, ARPES studies\cite{Valla-06,He-09NatPhys} show evidence of the usual nodal quasiparticles conventionally associated with a charge-2e $d$-wave SC.  Thus, if this is indeed a charge-4e SC, it is one analogous to that we have studied on the hexagonal lattice, although of course in LBCO the actual lattice structure is closer to that of a square lattice.

{\bf Acknowledgements:} Discussions with S. Raghu and E. Fradkin were helpful in clarifying the relevance of the Gross-Neveu-Ising universality class. We would like to thank the Supercomputer Center in Guangzhou for support. This work was supported in part by the NSFC under Grant No. 11474175 (YFJ, ZXL, and HY), and by NSF grant $\#$DMR 1265593 at Stanford (SAK).

\end{document}